\documentstyle[aps,multicol,psfig]{revtex}

\title{Reaction mechanisms for two-neutron halo breakup}
\author{E. Garrido}
\address{Instituto de Estructura de la Materia, CSIC, Serrano 123, E-28006
Madrid, Spain}
\author{D.V.~Fedorov, A.S.~Jensen and K. Riisager}
\address{Institute of Physics and Astronomy,
Aarhus University, DK-8000 Aarhus C, Denmark}
\begin{document}
\draft

\maketitle

\begin{abstract} 
We describe a reaction mechanism which is consistent with all
available experimental information of high energy three-body breakup
processes. The dominating channels are removal of one of the three
halo particles leaving the other two either undisturbed or absorbed.
We compare with the commonly used deceptive assumption of a decay
through two-body resonance states.  Our predictions can be tested by
measuring neutron-neutron invariant mass spectra.
\end{abstract}


\widetext
\tighten
\begin{multicols}{2}

\paragraph*{Introduction.}
 
Halo states are basically characterized as spatially extended weakly
bound systems.  Two-body halos interacting with a target present a
three-body scattering problem, which in practice is much more
complicated due to the intrinsic structure of the three constituent
particles. Three-body halos present analogously at least a four-body
problem, which has to be approximated preferentially by using physical
insights. Reactions for high beam energies allow a separation of the
degrees of freedom related to the fast relative projectile-target
coordinates and the slow intrinsic halo motion.

The properties of halo systems have been discussed intensely over the
last decade both in dripline nuclei \cite{rii94,han95,tan96,jon98} and
in molecular systems \cite{esr96,nie98}.  Precise definitions,
classification and occurrence conditions were recently attempted
\cite{rii00,jen00} although different from an earlier definition
\cite{goy95}.  Few-body concepts and techniques are successfully
applied in the descriptions \cite{gar98}, which to a large extent
focussed on nuclear three-body halos.  Much efforts have been devoted
to two-neutron Borromean halos like $^6$He (n+n+$^4$He) and $^{11}$Li
(n+n+$^{9}$Li) where two neutrons surround a core \cite{zhu93}. The
basic structure is essentially agreed upon while reaction descriptions
and analyses of measurements still are controversial.

Halo physics is a substantial part of experimental programs with
radioactive beams and it is urgent to root out widespread
misconceptions and clarify how the reactions proceed. Furthermore
these questions are of general interest as basic few-body reaction
problems.  Since the halo concept now is applied and exploited in
molecular physics we also may anticipate similar implications of
properly formulated reaction models.

The purpose of this letter is to (i) establish the reaction mechanism
for two-neutron halo breakup in high energy collisions with light
targets and in passing clarify the differences to the entirely
different mechanism due to the large charges of heavy targets, (ii)
investigate the validity of the erroneous but commonly used assumption
of breakup through resonances in the two-body subsystems. These
questions are crucial and answers urgently needed for understanding
reactions with halo nuclei.

\paragraph*{Reaction mechanisms.}

The dominating reaction channels for two-neutron halo breakup on light
targets are experimentally established \cite{ale98,aum99} and
theoretically described \cite{gar98,ber98} as removal of one
neutron or destruction of the core, thereby leaving the final state
with the core and the other neutron or with the two neutrons.

The decisive question in this context is which reaction mechanism is
responsible for the observed behavior? The reaction time for light
targets is short compared to the time scale of the intrinsic halo
motion.  For spatially extended systems the target can then remove one
of the halo particles instantaneously without disturbing the motion of
the other two particles. This means that the sudden approximation
basically is valid as accepted in several previous publications
\cite{ale98,aum99,zin97,chu97}.

The implication is that the remaining two-body system is left in its
initial state which, as unbound for Borromean systems, falls apart
influenced by the corresponding two-body interaction.  This decaying
two-body system is thus formed as a wave packet consisting of those
parts of the relative two-body wave function present within the
original three-body system, which precisely lead to the dominating
reaction products \cite{gar98}. The surviving wave packet then has a
large component describing the tail of the two-body wave function. The
short distance parts lead to a large extent to removal of more than
one particle at a time.  All other breakup reactions are analogously
described in this participant-spectator model (PSM).

\paragraph*{$R$-matrix formulation.}

The observed invariant two-body mass spectra and the momentum
distributions are routinely analyzed as arising from the decays of
low-lying two-body resonances or virtual $s$-states
\cite{ale98,aum99,zin97,chu97,sim98}.  These assumptions are in direct
contradiction to the short reaction time and the sudden
approximation. There is not sufficient time for the remaining two halo
particles to adjust their relative motion and populate corresponding
resonance states. This requires at least a reaction time comparable to
the intrinsic halo time scale.

Thus these analyses apparently invoke both the sudden approximation
and decay through resonances or virtual $s$-states.  These assumptions
are strictly incompatible except when these two-body states are
populated within the initial three-body system.  This is clearly seen
by constructing a Borromean system by adding a neutron to a
neutron-core resonance state.  The overlap of this and the real bound
state wave function may still be substantial, but rearrangements are
necessary to reach the bound three-body state, i.e. a novel few-body
system carrying otherwise inaccessible information about the off-shell
behavior of the nucleon-nucleon interaction.

The pertinent questions are what we can learn from measured two-body
invariant spectra and what information is in fact obtained by the
analyses using resonances or virtual $s$-states as intermediate
states.  The analyses, often erroneously claiming to use Breit-Wigner
shapes \cite{ale98,aum99,zin97,sim98}, are in fact based on $R$-matrix
theory \cite{lan58,mcv94}, where a complete basis of two-body
continuum states are used after removal of one of the three
particles. This basis could consist of ``correct'' low-lying resonance
states supplemented with a discretized or continuum higher lying set
of states.

In practice one proceeds by reducing the unspecified and unknown basis
to a few terms, i.e. usually one or two states. This reduction of
model space may be allowed if the basis consistently is renormalized,
i.e. the new basis includes properties of the excluded states. Thus
fitting in this context by use of a small basis seems to prohibit
interpretation in terms of the ``correct'' two-body resonance
states. In principle maintaining the basis without renormalization
would be correct but this presupposes exactly that knowledge about
these states, which is the very aim of the analyses. This problem
cannot be solved by increasing the employed model space until
convergence is reached and no renormalization is needed. A larger
space implies more parameters in the fitting procedure and
reproduction of the data is not unique.  The problem becomes
overdetermined and the extracted parameters inaccurate or directly
unreliable.

The analyses using two-body resonances or virtual $s$-states therefore
assume (i) that no renormalization due to truncation of model space is
needed, (ii) a reaction mechanism where only the ``clean'' two-body
resonance states or virtual $s$-states are populated and (iii) no
other (known or unexpected) reaction channel contribute. These
assumptions are at least inaccurate. The difficulties are enlarged
when more than one resonance or more than one reaction channel
contribute. When the assumptions are fairly well fulfilled the
interpretation would also be approximately correct.

\paragraph*{Computations.}

We shall concentrate on $^6$He and $^{11}$Li colliding with light
targets. We shall use the PSM formulation, where one halo particle
(the participant) interacts with the target while the other two halo
particles (the spectators) are left undisturbed \cite{gar98}. The
participant-target interaction is described by the phenomenological
optical model while the spectators are treated in the black sphere
approximation, i.e. they are absorbed within a given radius from the
target and otherwise they continue undisturbed.  This model faithfully
exploits the consequences of a short reaction time. We compute the
population of the two-body continuum states after the instantaneous
removal of the third particle. 

The $R$-matrix expressions of the invariant mass spectrum $d \sigma
/dE$ of the spectator system and the relative spectator momentum
distribution $P_{long}$ are \cite{ale98,zin97,sim98}
\begin{eqnarray}
  \frac{d \sigma}{dE} = \frac{\sigma_l}{2 \pi} \;
 \frac{ \Gamma(E)}{(E-E_{r})^2 + 
  0.25 \Gamma^2(E) }, \; 
 \Gamma=\Gamma_0 \frac{E^{l+0.5}}{E_r^{l+0.5}} \; ,
\label{e1}  \\
  P_{long}(p)  = \int^{\infty}_{E_{min}} \frac{d E} {\sqrt{E}} 
 \frac{d \sigma}{dE}, \hspace*{0.3cm} E_{min} = p^2/ 2 \mu
\label{e3}
\end{eqnarray}
where $\sigma_l$ is the total cross section, $l$ is the orbital
angular momentum, $\mu$ is the reduced mass, $E_r$ and $\Gamma_0$ are
position and width parameters.  The distributions are correlated
and should not be fitted independently.  Precisely the same procedure
applies both when the core and a neutron are the participant, i.e. the
final state consists of two neutrons or a neutron-core system,
respectively.  

The chosen observables amplify the effects of the assumed reaction
mechanism. We can then compare the experimental distributions both
with the PSM predictions \cite{gar98} and the $R$-matrix results
obtained by the decay through resonance assumption. This provides
evidence about the basic reaction mechanism.

\paragraph*{Neutron removal.} 

\begin{figure}
\centerline{\psfig{figure=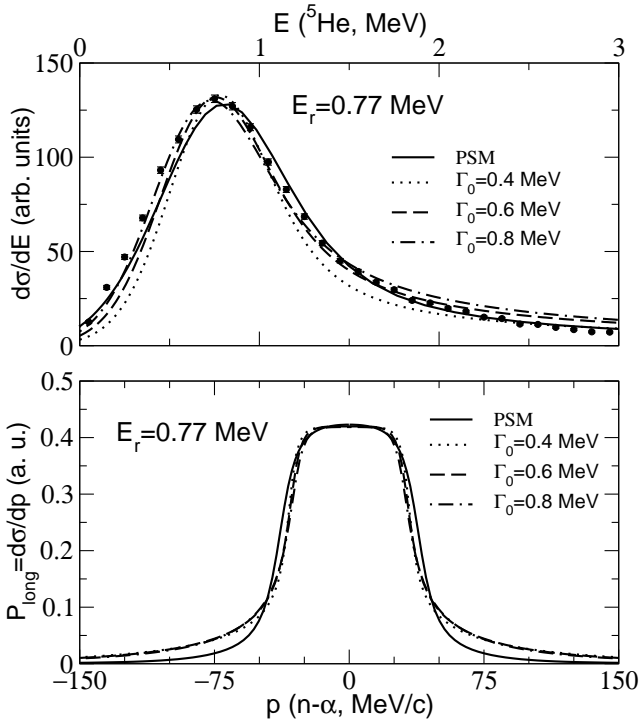,width=8.4cm,%
bbllx=0.5cm,bblly=3.6cm,bburx=19.2cm,bbury=24.6cm}}
\vspace*{0.3cm}
\caption[]{Neutron-$^4$He invariant mass spectrum (upper) and
longitudinal relative momentum distribution (lower) for breakup of 300
MeV/nucleon $^6$He projectile on a carbon target.  Experimental data
from \cite{ale98}. The solid curves are the PSM calculations
\cite{gar98} and the broken lines are obtained from Eqs.(\ref{e1}) and
(\ref{e3}) with $l=1$.  The invariant mass curves have been convoluted
with the instrumental response \cite{eml99}.}
\label{fig1}
\end{figure}

Absorption of one neutron from $^6$He produces a neutron-$^4$He
continuum state, which mainly is of $p_{3/2}$-character, since the
$p_{1/2}$ neutron-core state has a higher energy and the
$s_{1/2}$-wave is repulsive.  The corresponding invariant mass
spectrum and the relative momentum distribution are shown in
Fig.~\ref{fig1}.  The PSM computation agrees fairly well with the
measured invariant mass spectrum \cite{ale98}. The peak position
reflects the energy of 0.77 MeV of the neutron-$^4$He
$p_{3/2}$-resonance with the width of 0.5 MeV used in the PSM
computation.  Any value of $\Gamma_0$, see Eq.(\ref{e1}), from 0.4 MeV
to 0.8 MeV also reproduce the experiment fairly well.

For consistency the momentum distribution should now follow with the
same parameters. Indeed we see in Fig.~\ref{fig1} that the PSM results
resemble the (almost) width independent $R$-matrix fits confirming
that the width parameter for the n--$^4$He resonance can not be
determined in this way.  We also note the characteristic flat maximum
of $p$--waves.

\begin{figure}
\centerline{\psfig{figure=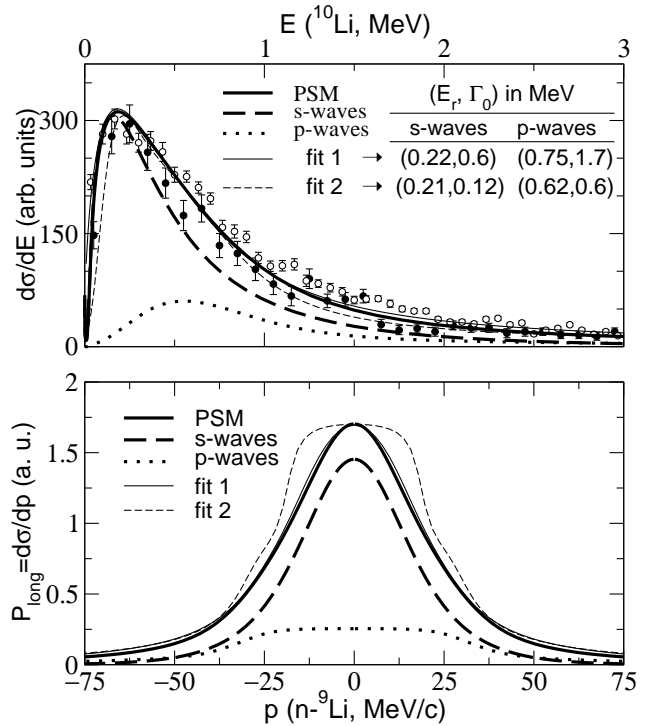,width=8.4cm,%
bbllx=0.5cm,bblly=3.6cm,bburx=19.2cm,bbury=24.6cm}}
\vspace*{0.3cm}
\caption[]{The same as Fig.~\ref{fig1} for neutron-$^9$Li for a
$^{11}$Li projectile. Both sets of experimental data
\cite{zin97,sim98} are normalized to the same maximum as the computed
spectrum. The thick solid curves are the PSM calculations consisting
of contributions from both neutron-core relative $s$ (thick dashed)
and $p$ (dotted) waves. The thin solid and dashed lines are weighted
averages of two distributions from Eqs.(\ref{e1}) and (\ref{e3})
corresponding to $s$ and $p$--resonances with parameters given in
MeV. }
\label{fig2}
\end{figure}

The $^{11}$Li system is different due to the core spin of $3/2$ and
the mixture of $s$ and $p$-waves in the subsystems.  The computed
three--body wave function contains around 60\% of $s$--wave and 40\%
of $p^2$--wave neutron-core configurations. The neutron-$^{9}$Li
system has a low lying virtual $s$--state at 240 keV and a
$p$--resonance at 0.5 MeV \cite{gar98}.  Neutron removal results in
the distributions shown in Fig.~\ref{fig2}.  The contribution to the
invariant mass spectrum from $s$-waves peaks at a very low energy
determined entirely from the phase space constraint and independent of
the position of the virtual $s$-state \cite{mcv94}. In contrast the
$p$-wave contribution peaks at the two-body resonance energy.  The
measured spectrum is fairly well reproduced by the PSM computation
again supporting the assumed initial three-body structure and the
reaction mechanism.

We compare in Fig.~\ref{fig2} with two different $R$-matrix fits.  In
the first the computed $s$ and $p$ contributions are fitted separately
thereby maintaining the same $p$--wave content.  In the second fit we
use the parameters in \cite{zin97}, which also reproduces rather well
the experimental (and PSM) data.  The $^{10}$Li structures underlying
these fits differ substantially as expressed clearly through the
different widths.  The $p$--wave contents also differ substantially,
i.e. about 35\% for the first and 70\% for the second fit
\cite{zin97}.

In the lower part of Fig.~\ref{fig2} we show the corresponding relative
momentum distribution. The PSM computation and the first fit produce a
very similar momentum distribution, while the second fit differs in
the central part due to the large fraction of $p$--waves that creates
the plateau at low relative momentum.  Therefore different fits of
invariant mass spectra of similar accuracy can produce rather
different momentum distributions due to emphasis of different features
of the distribution.

\paragraph*{Core destruction.}

The reaction assumptions can be tested by similar investigations of
the other important channel corresponding to destruction of the core
and leaving the two neutrons as spectators. The final states then
consists of two neutrons for both $^6$He and $^{11}$Li. Thus we can
separate effects of initial and final state structures.  However, this
assumes that fragments from the core interacting strongly with
neutrons are excluded from the data.  In the PSM computations,
discussed in connection with Figs.~\ref{fig1} and \ref{fig2}, the
spectra are sensitive to the initial three--body structure. The root
mean square distance between the neutrons is more than 6 fm for
$^{11}$Li and less than 4.5 fm for $^6$He.  The neutron--neutron
invariant mass spectrum and the corresponding momentum distribution
are then both expected to be substantially narrower for $^{11}$Li than
for $^6$He.  The same consistent PSM model has been tested on many
other, relative and absolute values of observables for neutron removal
and core breakup reactions for both projectiles \cite{gar98}. The
agreement with available experimental data is overall very convincing.

\begin{figure}
\centerline{\psfig{figure=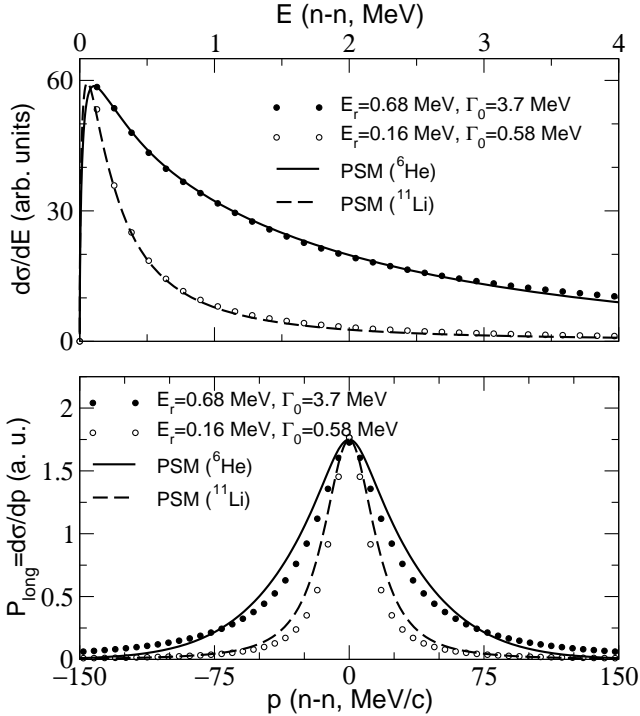,width=8.4cm,%
bbllx=0.5cm,bblly=3.6cm,bburx=19.2cm,bbury=24.6cm}}
\vspace*{0.3cm}
\caption[]{The same as Fig.~\ref{fig1} for the neutron-neutron system
for $^6$He and $^{11}$Li projectiles. The points are the two
$R$-matrix fits given in the figure.}
\label{fig3}
\end{figure}

Further tests of the PSM model (and the $R$-matrix analyses) would be
measurements comparing to the predictions presented in
Fig.~\ref{fig3}.  The neutron-neutron relative $s$-waves are
completely dominating for both $^6$He and $^{11}$Li and consequently
the invariant mass spectra have very low-lying peaks. Both spectra and
momentum distributions are qualitatively similar for the two cases,
but quantitatively the $^{11}$Li results are much narrower than those
of $^6$He.  The $R$-matrix distributions fitting the two PSM curves in
Fig.~\ref{fig3} correspond to very different energy and width
parameters without any connection to the known neutron-neutron
scattering properties.  The PSM model predicts different
neutron-neutron spectra for $^6$He and $^{11}$Li after core breakup. A
reaction mechanism populating final state two-body resonances
independent of the initial structure must predict identical
neutron-neutron invariant mass spectra for both
projectiles. Experimental data could distinguish between these models.

\paragraph*{Conclusions.}

The dominating reaction channels for high energy breakup of Borromean
three-body halos on light targets are one-particle removal and
subsequent decays of the wave packets created in these processes. The
reaction time is short and any resonance structure of the remaining
two-body system is populated with the amount already present in the
initial three-body wave function. All available experimental data for
high energy three-body breakup on light nuclei are consistent with
this reaction mechanism. For heavy targets the reaction mechanism for
the dominating channel is quite different proceeding through a gentle
excitation of the three-body continuum by the Coulomb interaction.

Analyses assuming instantaneous removal of either a neutron or the
core, while populating resonances in the remaining two-body system,
are conceptually inconsistent. An invariant mass spectrum reproducing
the data only reflects that the corresponding energy distribution was
present immediately after the final state two-body system was
isolated.  The problem is especially enlarged when more than one
resonance or virtual state are important for the two-body
subsystems. The inconsistency is highlighted in spectra obtained after
core breakup, where the final states are identical (two neutrons).
Then the resulting distributions should also be identical for
different two-neutron halo projectiles, even when the initial
three-body structure differs substantially.  This is in clear
disagreement with elaborated consistent model computations reproducing
essentially all available data.  In any case, the neutron--neutron
invariant mass spectra provide direct evidence of the breakup reaction
mechanism.

We conclude that correct interpretation of the invariant spectra
almost inevitably require a consistent model, i.e. the three-body
structure, the two-body interactions and the reaction mechanisms must
be (approximately) correct. Even then interplay between the different
ingredients may produce misleading results. The $R$-matrix analyses
rely on assumptions or computations of the initial steps producing an
isolated two-body system.

\paragraph*{Acknowledgement.}  We thank L.V. Chulkov and H. Emling for
illuminating discussions.

\end{multicols}


\begin{thebibliography}{99}

\bibitem{rii94} K.~Riisager, Rev. Mod. Phys. {\bf 66}, 1105 (1994).

\bibitem{han95} P.G. Hansen, A.S. Jensen and B. Jonson, 
Ann. Rev. Nucl. Part. Sci. {\bf 45}, 591 (1995). 

\bibitem{tan96} I. Tanihata, J. Phys.  {\bf G22}, 157 (1996). 

\bibitem{jon98} B. Jonson and K. Riisager, Phil. Trans. R. Soc. Lond.
{\bf A356}, 2063 (1998).

\bibitem{esr96} B.D.~Esry, C.D.~Lin and C.H.~Greene, Phys.~Rev. {\bf A54},
394 (1996).

\bibitem{nie98} E.~Nielsen, D.V. Fedorov and A.S. Jensen,
J.~Phys. {\bf B31}, 4085 (1998).

\bibitem{rii00} K. Riisager, D.V. Fedorov and A.S. Jensen,
Europhysics Lett. {\bf 49}, 547 (2000).

\bibitem{jen00} A.S. Jensen and K. Riisager,
Phys. Lett.  {\bf B 480}, 39 (2000). 

\bibitem{goy95} J.~Goy, J.-M.~Richard and S.~Fleck,
Phys.~Rev. {\bf A52}, 3511 (1995).

\bibitem{gar98} E. Garrido, D.V. Fedorov and A.S. Jensen,
Europhysics Lett. {\bf 43}, 735 (1998);  {\bf 50}, 735 (2000); 
Phys. Rev. {\bf C59}, 1272 (1999); Phys. Lett.  {\bf B 480}, 32 (2000). 

\bibitem{zhu93} M.V. Zhukov, B.V. Danilin, D.V. Fedorov, J.M. Bang,
I.J. Thompson and J.S. Vaagen, Phys. Rep. {\bf 231}, 151 (1993).

\bibitem{ale98} D. Aleksandrov {\it et al.}, Nucl. Phys. 
{\bf A633}, 234 (1998).

\bibitem{aum99} T. Aumann {\it et al.}, Phys. Rev. {\bf C59}, 1252 (1999).

\bibitem{ber98} G.F. Bertsch, K. Hencken and H. Esbensen,
Phys. Rev. {\bf C57}, 1366 (1998).

\bibitem{zin97} M. Zinser {\it et al.}, Nucl. Phys. {\bf A619}, 151
(1997).

\bibitem{chu97} L. Chulkov and G. Schrieder, Z. Phys. {\bf A359}, 231 (1997).

\bibitem{sim98} H. Simon, Ph.D. Thesis, Technische Universit\"{a}t, 
Darmstadt (1998).

\bibitem{lan58} A.M. Lane and R.G. Thomas, 
Rev. Mod. Phys. {\bf 30}, 257 (1958).

\bibitem{mcv94} K.W. McVoy and P. Isacker, Nucl. Phys. {\bf A576}, 157 (1994).

\bibitem{eml99} H. Emling, Private communication.


\end{thebibliography}
\end{document}